\documentclass[aps,prl,twocolumn,showpacs,amsmath,amssymb,superscriptaddress]{revtex4}
\usepackage{graphicx}
\usepackage{amssymb}
\usepackage{natbib}
\usepackage{color}

\begin{document}

\title{Anisotropic neutron spin resonance in underdoped  superconducting NaFe$_{1-x}$Co$_x$As}
\author{Chenglin Zhang}
\thanks{These authors made equal contributions to this paper}
\affiliation{Department of Physics and Astronomy, Rice University, Houston, Texas 77005, USA}
\affiliation{ Department of Physics and Astronomy,
The University of Tennessee, Knoxville, Tennessee 37996-1200, USA }
\author{Yu Song}
\thanks{These authors made equal contributions to this paper}
\affiliation{Department of Physics and Astronomy, Rice University, Houston, Texas 77005, USA}
\author{L.-P. Regnault}
\affiliation{SPSMS-MDN, UMR-E CEA/UJF-Grenoble 1, INAC, Grenoble, F-38054, France}
\author{Yixi Su}
\affiliation{
J\"{u}lich Centre for Neutron Science, Forschungszentrum J\"{u}lich GmbH, 
Outstation at MLZ, D-85747 Garching, Germany
}
\author{M. Enderle}
\affiliation{Institut Laue-Langevin, 6, rue Jules Horowitz, BP 156, 38042 Grenoble Cedex 9, France}
\author{J. Kulda}
\affiliation{Institut Laue-Langevin, 6, rue Jules Horowitz, BP 156, 38042 Grenoble Cedex 9, France}
\author{Guotai Tan}
\affiliation{ Department of Physics and Astronomy,
The University of Tennessee, Knoxville, Tennessee 37996-1200, USA }
\author{Zachary C. Sims}
\affiliation{ Department of Physics and Astronomy,
The University of Tennessee, Knoxville, Tennessee 37996-1200, USA }
\author{Takeshi Egami}
\affiliation{ Department of Physics and Astronomy,
The University of Tennessee, Knoxville, Tennessee 37996-1200, USA }
\affiliation{Oak Ridge National Laboratory, Oak Ridge, Tennessee 37831, USA}
\author{Qimiao Si}
\affiliation{Department of Physics and Astronomy, Rice University, Houston, Texas 77005, USA}
\author{Pengcheng Dai}
\email{pdai@rice.edu}
\affiliation{Department of Physics and Astronomy, Rice University, Houston, Texas 77005, USA}

\date{\today}

\begin{abstract}
We use polarized inelastic neutron scattering (INS) to study spin excitations in 
superconducting NaFe$_{0.985}$Co$_{0.015}$As (C15) with static antiferromagnetic (AF) 
order along the $a$-axis of the orthorhombic structure and NaFe$_{0.935}$Co$_{0.045}$As (C45) without AF order.
In previous unpolarized INS work, spin excitations in C15 were found to have 
a dispersive sharp resonance near $E_{r1}=3.25$ meV and a broad dispersionless mode at $E_{r2}=6$ meV.
Our neutron polarization analysis reveals that the dispersive resonance in C15 
is highly anisotropic and polarized along the $a$- and $c$-axis, 
while the dispersionless mode is isotropic similar to that of C45.
Since the $a$-axis polarized spin excitations 
of the anisotropic resonance appear below $T_c$, our data suggests that the itinerant electrons contributing to the magnetism 
are also coupled to the superconductivity.
\end{abstract}

\maketitle

Superconductivity in iron pnictides occurs when the antiferromagnetic (AF) order in their parent compounds is 
suppressed via electron or hole-doping \cite{kamihara,cruz,cwchu,slli,ysong,johnston,dai}.  In the undoped state, iron pnictides 
exhibit a tetragonal to orthorhombic lattice distortion at temperature $T_s$, followed by a
paramagnetic to AF phase transition at
$T_N$ with a collinear AF structure and the ordered moments along the $a$-axis
of the orthorhombic lattice [inset in Fig. 1(a) or $M_a$]  \cite{kamihara,cruz,cwchu,slli,ysong,johnston,dai}.
Upon doping to induce superconductivity, the most prominent feature in the spin excitations spectrum is a 
neutron spin resonance arising below $T_c$ at the in-plane AF ordering wave vector ${\bf Q_{AF}}=(1,0)$ \cite{Christianson08,mdlumsden09,schi09,dsinosov10,chlee13,mgkim13,dsinosov11}.  For hole and electron-doped BaFe$_2$As$_2$ family of materials \cite{Christianson08,mdlumsden09,schi09,dsinosov10,chlee13,mgkim13,dsinosov11}, the resonance occurs at an energy $E$ believed to be associated with the superconducting gap energies at the hole and electron Fermi surfaces near $\Gamma$
and $M$ points in the reciprocal space, respectively \cite{Hirschfeld}. In the case of electron-doped superconducting NaFe$_{1-x}$Co$_x$As [Fig. 1(a)] \cite{parker,afwang12,gttan}, unpolarized inelastic neutron scattering (INS) experiments reveal that superconductivity 
induces a dispersive sharp resonance near $E_{r1}=3.25$ meV and a broad dispersionless
mode at $E_{r2}=6$ meV at ${\bf Q_{AF}}=(1,0)$ in the underdoped NaFe$_{0.985}$Co$_{0.015}$As (C15) with static AF 
order ($T_c=15$ K and $T_N=30$ K) \cite{clzhangprl}, while only a single resonance at $E_r=7$ meV in the overdoped NaFe$_{0.935}$Co$_{0.045}$As (C45, $T_c=18$ K) \cite{clzhangprb}.

The presence of double resonance in superconducting C15 coexisting with 
static AF order \cite{clzhangprl} has inspired much discussion on its microscopic origin.
In one class of models, the double resonance arises from superconductivity coexisting with
static AF order \cite{rowe,weicheng}.  In this picture, through an averaging effect in twinned samples, 
the double resonances observed at, say, ${\bf Q}=(1,0)$, are interpreted as 
reflecting one single resonance at the 
AF zone center ${\bf Q_{AF}}=(1,0)$ and one at the wave vector ${\bf Q}^{\prime}=(0,1)$ \cite{rowe,weicheng}.
Alternatively, the double resonance in C15 may probe the superconducting gap
anisotropy in the underdoped regime seen in the angle resolved photoemission experiments \cite{clzhangprl,qqge13,ryu14}. Here, the
orbital-selective pairing gives rise to gap anisotropy along a
Fermi surface with hybridized orbital characters, resulting a split of the neutron spin 
resonance \cite{ryu14}.  Since the resonance is generally believed to result from 
a triplet excitation of the singlet electron Cooper pairs associated with 
isotropic paramagnetic spin excitations [$M_a=M_b=M_c$ in the inset of Fig. 1(a)] \cite{eschrig}, 
a determination of its spatial anisotropy is important for understanding the double resonance and 
its microscopic origin.

In this paper, we report polarized INS studies of undoped C15 and overdoped 
C45 \cite{clzhangprl,clzhangprb}. 
We find that the dispersive resonance in C15  
is highly anisotropic and polarized along the $a$- and $c$-axis ($M_a,M_c>0$) with no contribution from the $b$-axis ($M_b=0$).
However, the dispersionless resonances in C15 and C45 are isotropic with ($M_a=M_b=M_c$), consistent with the singlet-to-triplet excitations \cite{eschrig}.  
Since spin waves in the undoped NaFeAs are entirely $c$-axis polarized for energies below $\sim$10 meV \cite{ysong13b}, 
the appearance of $a$-axis (longitudinally) polarized 
resonance in the AF ordered C15 below $T_c$  
indicates that the dispersive resonance is unlikely to arise from coexisting AF order with superconductivity \cite{rowe,weicheng}.
 Instead, the data is consistent with the orbit-selective superconducting gap anisotropy \cite{ryu14}, suggesting that the itinerant electron contributions to the magnetism, revealed as longitudinal spin excitations in undoped parent compounds \cite{cwang}, are also coupled to superconductivity.

\begin{figure}[t]
\includegraphics[scale=.45]{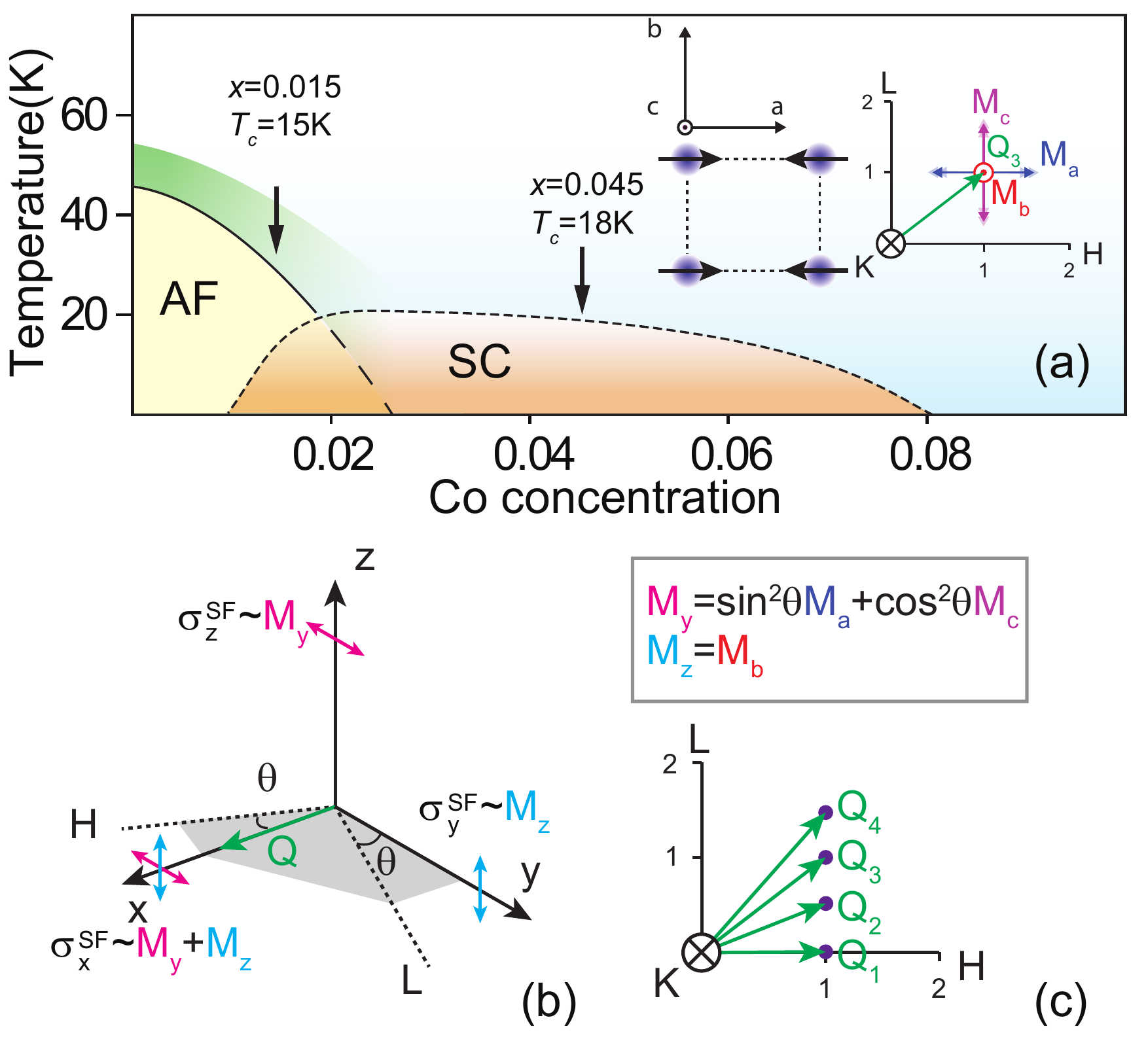}
\caption{
(Color online) (a) The phase diagram of NaFe$_{1-x}$Co$_x$As with $x=0.015,0.045$ marked by vertical arrows \cite{parker}.
The left inset shows the orthorhombic unit cell of NaFeAs with 
the arrows indicating directions of the ordered moments.  The right inset shows reciprocal space in the $[H,0,L]$ scattering
plane.  The blue, red, and purple arrows mark spin excitations along the $M_a$, $M_b$, and $M_c$ directions, respectively.
(b) The relationship between the neutron polarization directions ($x,y,z$) and the probed reciprocal space.
  The angle between $x$-direction and the $H$-axis is denoted as $\theta$.  
$\sigma_{x}^{\rm SF}$ contains both $M_y$ and $M_z$ magnetic components, whereas only $M_y$ and $M_z$ 
contribute to $\sigma_{z}^{\rm SF}$ and $\sigma_{y}^{\rm SF}$, respectively. 
(c) The ${\bf Q_1}=(1,0,0)$,  ${\bf Q_2}=(1,0,0.5)$, ${\bf Q_3}=(1,0,1)$, ${\bf Q_4}=(1,0,1.5)$ mark the
probed reciprocal space. 
}
\end{figure}

The inset in Fig. 1(a) shows the AF structure of NaFeAs with orthorhombic lattice parameters 
$a=5.589$, $b= 5.569$ and $c=6.991$ \AA\ \cite{slli}.
We define momentum transfer ${\bf Q}$ in three-dimensional 
reciprocal space in \AA$^{-1}$ as $\textbf{Q}=H\textbf{a}^\ast+K\textbf{b}^\ast+L\textbf{c}^\ast$,
where
$H$, $K$, and $L$ are Miller indices and
${\bf a}^\ast=\hat{{\bf a}}2\pi/a$, ${\bf b}^\ast=\hat{{\bf b}}2\pi/b$, ${\bf c}^\ast=\hat{{\bf c}}2\pi/c$.
  In this notation,
the AF Bragg peaks and zone centers occur at $[1,0,L]$ with $L=0.5,1.5,\cdots$, while the AF zone boundaries along the $c$-axis occur at $L=0,1,2,\cdots$ \cite{slli}.
The dynamic susceptibility along the $a$-, $b$-, and $c$-axis directions corrected for the Bose population factor are 
marked as $M_a$, $M_b$, and $M_c$, respectively [inset in Fig. 1(a)] \cite{olly}.  Our polarized INS experiments were
carried out using the IN20 and IN22 triple-axis spectrometers at the Institut
Laue-Langevin, Grenoble, France \cite{ysong13b,cwang,olly,boothroyd,prokes,hqluo13}.  
Single crystals of C14 and C45 used in previous unpolarized neutron scattering experiments are used in 
the present experiment \cite{clzhangprl,clzhangprb}.
The quality of our single crystals of NaFe$_{1-x}$Co$_x$As has been reported in 
previous heat capacity \cite{gttan}, angle resolved photoemission spectroscopy \cite{qqge13},
and nuclear magnetic resonance \cite{soh,soh13b} experiments. 
We define the neutron polarization directions
along ${\bf Q}$ as $x$, perpendicular to ${\bf Q}$ but in the scattering plane
as $y$, and perpendicular to ${\bf Q}$ and the scattering plane as $z$, respectively [Figs. 1(b)].  At wave vector ${\bf Q}$, 
 one can probe magnetic responses within the $y-z$ plane ($M_y$ and $M_z$), giving $M_y=M_a\sin^2\theta+M_c\cos^2\theta$ and $M_z=M_b$,
where the angle between ${\bf Q}$ and $[H,0,0]$ is $\theta$ [Fig. 1(b)] \cite{olly}.  By probing two or more 
equivalent AF wave vectors with different
angle $\theta$, we can conclusively determine $M_a$, $M_b$, and $M_c$ (Fig. 1c) \cite{ysong13b}. 

\begin{figure}[t]
\includegraphics[scale=.45]{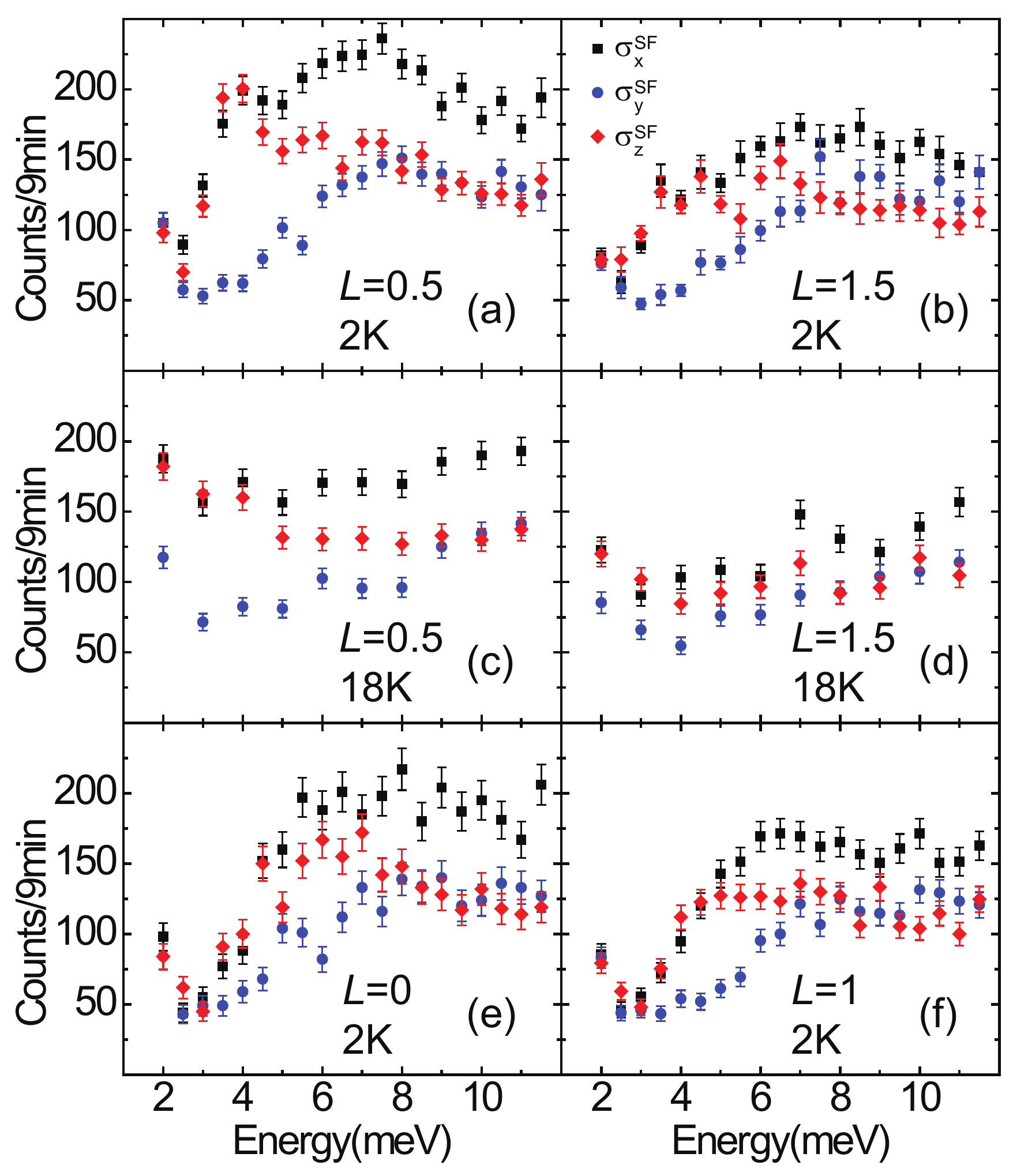}
\caption{
(Color online)  Neutron spin-flip scattering $\sigma_x^{\rm SF}$, $\sigma_y^{\rm SF}$, and $\sigma_z^{\rm SF}$
at (a) ${\bf Q_2}$ (b) ${\bf Q_4}$ in the superconducting state. (c,d) Identical scans in the normal state.
(e,f) $\sigma_x^{\rm SF}$, $\sigma_y^{\rm SF}$, and $\sigma_z^{\rm SF}$ at ${\bf Q_1}$ and ${\bf Q_3}$, respectively.
}
\end{figure}

To establish the spin excitation
anisotropy in NaFe$_{1-x}$Co$_{x}$As, we carried out polarized INS in C15 \cite{clzhangprl}.
Figures 2(a) and 2(b) show neutron spin-flip (SF) scattering for neutron polarizations along the 
$x$ ($\sigma_x^{\rm SF}$), $y$ ($\sigma_y^{\rm SF}$), and $z$ ($\sigma_z^{\rm SF}$) directions at the AF zone centers ${\bf Q_2}={\bf Q_{AF}}=(1,0,0.5)$ and 
${\bf Q_4}=(1,0,1.5)$, respectively, in the superconducting state ($T=2$ K).  
We find that $\sigma_x^{\rm SF}$ has a narrow peak at $E_{r1}\approx 4$ meV and a broad peak at 
 $E_{r2}=7$ meV, consistent with the two resonances in previous unpolarized work \cite{clzhangprl}.
However, the situation is rather different for $\sigma_y^{\rm SF}\sim M_z=M_b$ and 
$\sigma_z^{\rm SF}\sim M_y$.  While $\sigma_z^{\rm SF}$ has clear peaks at $E_{r1}\approx 4$ and $E_{r2}=7$ meV,
$\sigma_y^{\rm SF}$ has a broad peak at $E_{r2}=7$ meV and
is featureless at $E_{r1}\approx 4$ meV. 
Identical scans in the normal state 
($T=18$ K) reveal magnetic 
 anisotropy below $8$ meV with $\sigma_x^{\rm SF}\geq \sigma_z^{\rm SF} >\sigma_y^{\rm SF}$ [Figs. 2(c) and 2(d)].
Figures 2(e) and 2(f) show similar data
 at the AF zone boundaries ${\bf Q_1}=(1,0,0)$ and 
${\bf Q_3}=(1,0,1)$, respectively, in the superconducting state.

\begin{figure}[t]
\includegraphics[scale=.35]{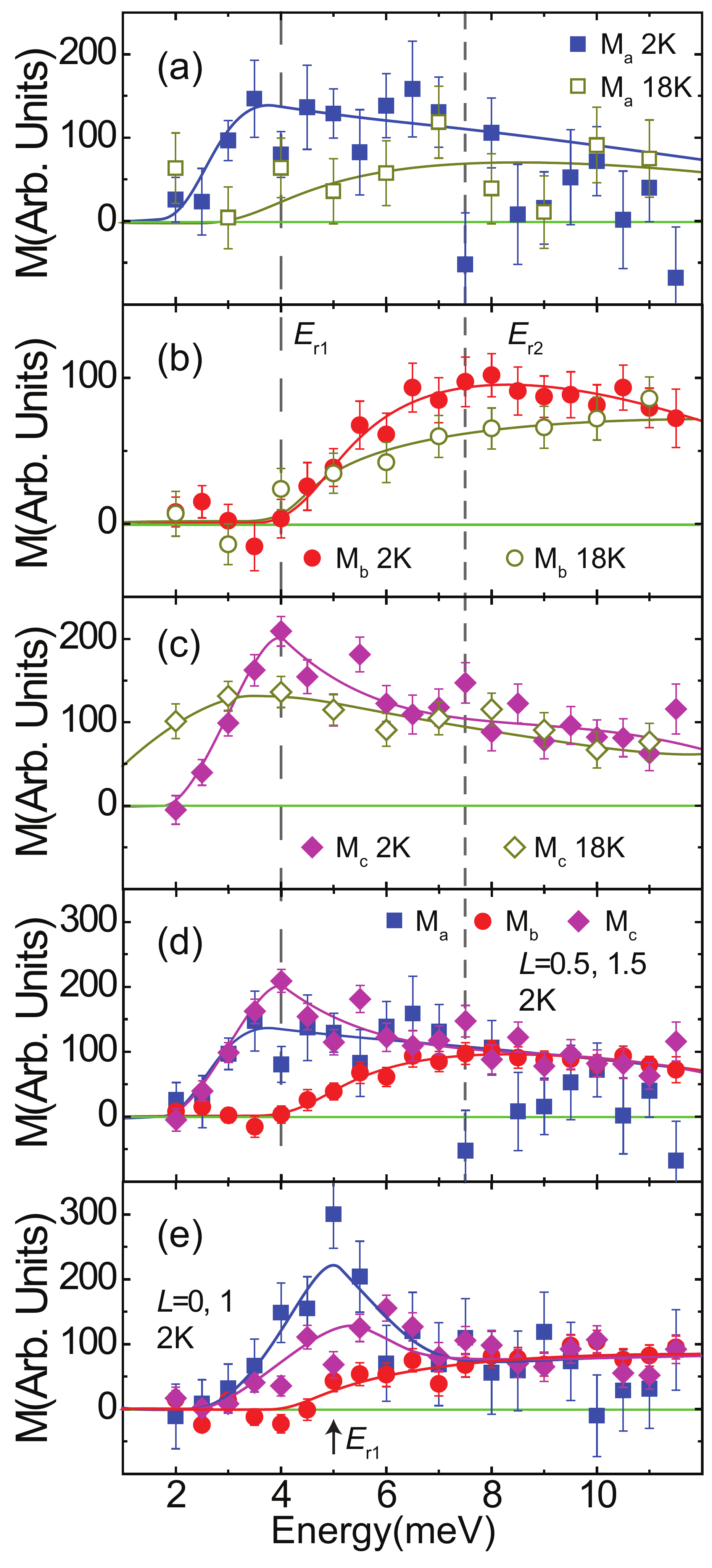}
\caption{
(Color online) Energy dependence of the dynamic susceptibility (a) $M_a$, (b) $M_b$, and (c) $M_c$ above and below $T_c$.
Energy dependence of $M_a$, $M_b$, and $M_c$ in the superconducting state at (d) AF zone center and (e) zone boundary.
The vertical dashed lines indicate energies of $E_{r1}$ and $E_{r2}$, and the solid lines are guides to the eye.
 }
 \end{figure}

\begin{figure}[t]
\includegraphics[scale=.45]{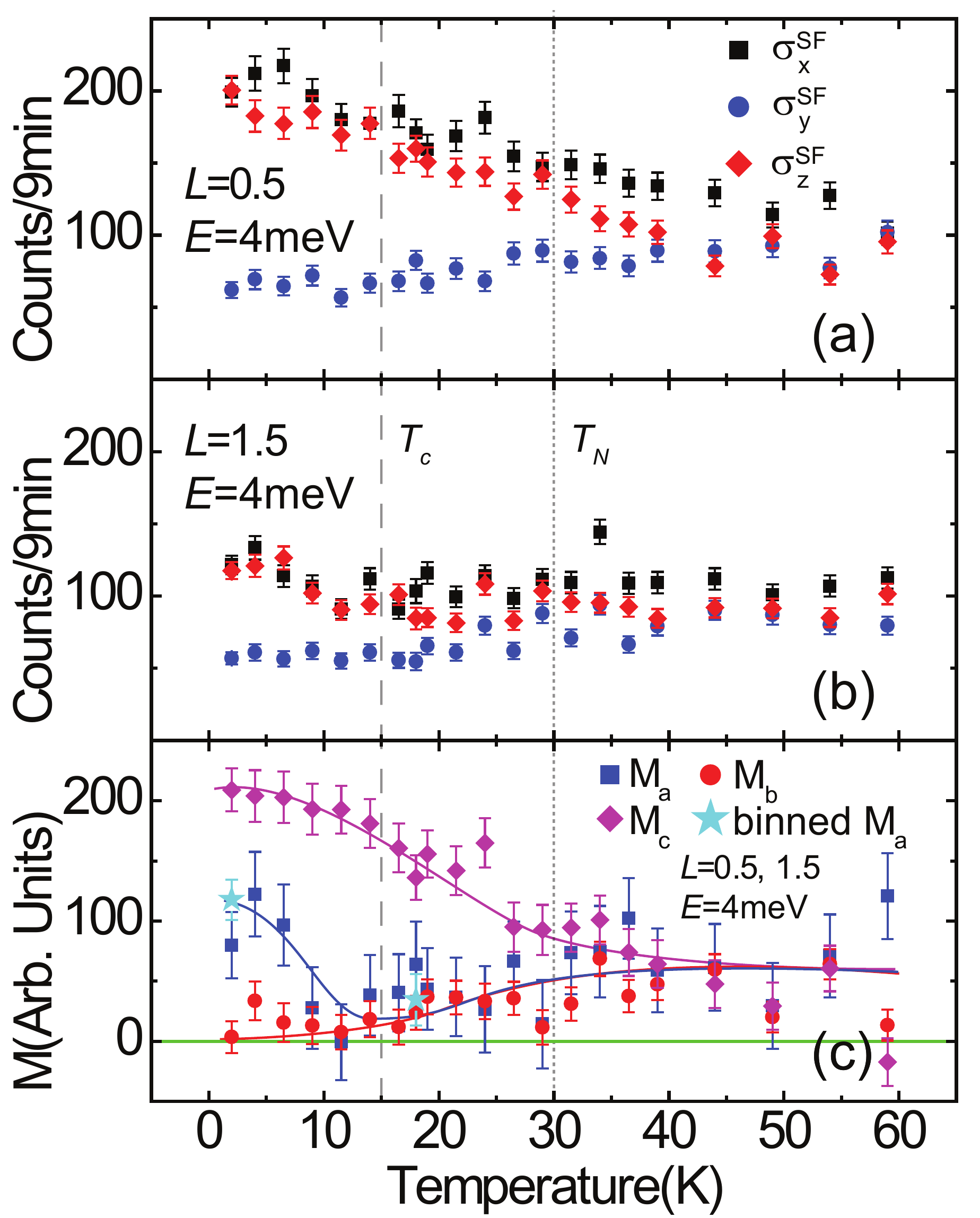}
\caption{
(Color online) Temperature dependence of  $\sigma_x^{SF}$, $\sigma_y^{SF}$, and $\sigma_z^{SF}$
at $E=4$ meV and (a) ${\bf Q_{AF}}={\bf Q_2}=(1,0,0.5)$, (b) ${\bf Q_{AF}}={\bf Q_4}=(1,0,1.5)$. 
(c) Temperature dependence of the estimated $M_a$, $M_b$, and $M_c$ at $E=4$ meV. 
The stars below and above $T_c$ are from binned energy scan data near $E=4$ meV.
The vertical dashed lines mark $T_N$ and
$T_c$.  The solid lines are guides to the eye.
 }
\end{figure}

Using data in Fig. 2, we determine the magnetic anisotropy 
$M_a$, $M_b$, and $M_c$ in C15 \cite{suppl}. Figures 3(a), 3(b), and 3(c) show 
energy dependence of the $M_a$, $M_b$, and $M_c$, respectively, above and below $T_c$.
While spin excitations along the $M_a$ and $M_c$ directions show clear peaks in the 
superconducting state above the normal state scattering near $E_{r1}\approx 4$ meV [Figs. 3(a) and 3(c)], 
there are no detectable difference in 
$M_b$ across $T_c$ at $E_{r1}\approx 4$ meV [Fig. 3(b)].  In contrast, spin excitations along the $b$-axis direction
($M_b$) show the most dramatic change below $T_c$ near $E_{r2}\approx 7$ meV.
Figure 3(d) plots the energy dependence of the $M_a$, $M_b$, and $M_c$ in the superconducting state, showing 
a large spin gap below $\sim$4 meV and isotropic paramagnetic scattering above $\sim$7 meV.  Therefore, the resonance 
at $E_{r1}\approx 4$ meV is composed of spin excitations polarized along the $a$ ($M_a$) and $c$ ($M_c$) axes, while the mode at $E_{r2}=7$ meV 
is isotropic in space with $M_a\approx M_b\approx M_c$.  
Figure 3(e) shows the energy dependence of $M_a$, $M_b$, and $M_c$ in the superconducting state
at the AF zone boundary ${\bf Q_1}=(1,0,0)$.  Consistent with unpolarized INS work \cite{clzhangprl}, 
we find that the resonance at $E_{r1}\approx 4$ meV shifted up in energy to $E_{r1}\approx 5$ meV while the broad
resonance remains unchanged at $E_{r2}\approx 7$ meV.  Similar to the data at the AF zone center, the resonance 
at $E_{r1}\approx 5$ meV has $M_a$ and $M_c$ components with $M_b=0$ and spin excitations are isotropic for 
energies above 7 meV.

Since the resonance 
at the AF zone center shows clear magnetic anisotropy in the superconducting state, 
we carried out temperature dependent measurements of the spin-flip scattering
at $E_{r1}\approx 4$ meV.  Figure 4(a) and 4(b) shows  
the $\sigma_x^{\rm SF}$, $\sigma_y^{\rm SF}$, and $\sigma_z^{\rm SF}$ scattering at
the AF wave vectors ${\bf Q_2}=(1,0,0.5)$ and 
${\bf Q_4}=(1,0,1.5)$, respectively.  In previous unpolarized measurements, temperature
dependence of the scattering at $E_{r1}=3.25$ meV reveals a kink at $T_N\approx 30$ K 
and a clear enhancement below $T_c=15$ K \cite{clzhangprl}. Figure 4(c) shows temperature dependence of
the $M_a$, $M_b$ and $M_c$ obtained by using data in Figs. 4(a) and 4(b). In the paramagnetic state,
spin excitations are isotropic with $M_a=M_b=M_c$.  On cooling to below $T_N$,
spin excitations are dominated by $c$-axis polarized moment 
$M_c$ with rapid suppression of $M_a$ and $M_b$.  
These results are consistent with previous polarized INS
work that show entirely $c$-axis polarized low-energy spin waves ($M_c$) in the AF ordered state of NaFeAs \cite{ysong13b}. 
While $M_c$ continues to increase with decreasing temperature and shows no obvious 
anomaly across $T_c$, $M_a$ increases dramatically below $T_c$ similar to the superconducting order parameter.
In contrast, there are no $b$-axis polarized magnetic scattering below $T_c$ ($M_b=0$).

Having established the spin excitation anisotropy in electron 
underdoped C15 with static AF order and double resonances \cite{clzhangprl}, it would be 
interesting to determine what happens in electron-overdoped C45, which has no static AF order 
and a sharp resonance at $E_r=7$ meV \cite{clzhangprb}. 
From polarized INS data presented in \cite{suppl}, we conclude that magnetic scattering in electron overdoped C45 
is isotropic at all energies and temperatures with
$M_a=M_b=M_c$.

In previous polarized INS work on the parent compound BaFe$_2$As$_2$, spin waves below $\sim$12 meV are 
entirely $c$-axis polarized ($M_c$) \cite{qureshi,cwang}. Upon electron-doping to BaFe$_2$As$_2$ via Co 
substitution to induce optimal 
superconductivity, polarized INS found evidence 
for two neutron spin resonance-like excitations with one isotropic mode ($M_a=M_b=M_c$) at an energy of
$\sim$8 meV and a purely $c$-axis polarized mode ($M_c$) at $\sim$4 meV \cite{steffens}. 
These results suggest that the low-energy $c$-axis polarized mode arises from the $c$-axis polarized 
spin waves in the parent compound \cite{steffens}.
The discovery that the $E_{r1}=3.25$ meV sharp resonance in the underdoped Co15 
 is composed of a primarily longitudinally polarized mode coupled to superconductivity is clearly different from 
those of electron-doped BaFe$_2$As$_2$ superconductors \cite{hqluo13,steffens}. 
Since the 
$c$-axis polarized excitations of the $E_{r1}=3.25$ meV resonance
 shows no anomaly across $T_c$ in the C15 [$M_c$ in Fig. 4(c)] and its  
spin anisotropy features at the AF zone boundary
along the $c$-axis [$L=0,1$, Fig.~3(e)] are rather similar to those at the zone center [$L=0.5,1.5$, Fig.~3(d)], 
the mode is unlikely to arise from the $c$-axis polarized spin waves coupling with 
superconductivity \cite{rowe,weicheng}. 
By contrast, these observations are consistent with the
scenario based on orbit-selectivity-induced superconducting gap anisotropy \cite{ryu14}. 
In this latter picture, the anisotropy of the spin
resonance arises from a spin-orbit coupling, which operates when the resonance comes 
from the superconducting quasiparticle-quasihole excitations that are associated with
 both the $3d$ $xy$ and $xz/yz$ orbitals \cite{ryu14}.

To summarize, our discovery of a primarily longitudinally polarized resonance 
at $E_{r1}\approx 4$ meV implies that the longitudinal
spin excitations typically associated with magnetism from itinerant electrons 
are also coupled to superconductivity. This suggests that itinerant electrons 
play an important role in the low-energy spin dynamics of the superconducting state.
At the same time, our study provides further evidence for the role of electron-correlation-induced orbital
selectivity in the superconducting state, underscoring the importance
of the local correlations to the superconducting resonance excitations.
Taken together, our results indicate that even the nominally itinerant-electron
contributions to the low-energy spin excitations encode the effects of local electronic correlations. 
This is a new insight in the microscopic physics of the iron-pnictide superconductors.

We thank useful discussions with Weicheng Lv, ilya Eremin and Rong Yu.
The crystal growth and neutron scattering work 
at Rice/UTK was supported by the US DOE, BES, DE-FG02-05ER46202 (P.D.). 
Work at Rice University was supported by the NSF Grant No. DMR-1309531 (Q.S.)
and the Robert A. Welch Foundation Grant Nos. C-1839 (P.D.) and C-1411 (Q.S.).
C.L.Z and T.E are partially supported by the US DOE BES through the EPSCoR grant, DE-FG02-08ER46528.

\end{document}